\documentclass[namedreferences]{solarphysics}

\usepackage[hyperref,optionalrh,natbib]{spr-sola-addons} % For Solar Physics 
\usepackage{graphicx}        % For eps figures, newer & more powerfull
\usepackage{color}           % For color text: \color command
\usepackage{url}       
\usepackage{breakurl}        % For breaking URLs easily trough lines
\usepackage{rotating}
\newcommand{\aap}{    {\it Astron. Astrophys.}}

\newcommand{\apj}{    {\it Astrophys. J.}}
\newcommand{\apjl}{   {\it Astrophys. J. Lett.}}

\newcommand{\jastp}{  {\it J. Atmos. Solar-Terr. Phys.}} 
\newcommand{\jgr}{    {\it J. Geophys. Res. (Space Phys.)}}

\newcommand{\planss}{   {\it Planet. Space Sci.}}

\newcommand{\solphys}{{\it Solar Phys.}}
 
\newcommand{\ssr}{    {\it Space Sci. Rev.}}

\begin{document}

\begin{article}

\begin{opening}

\title{On the Statistical Relationship between CME Speed and Soft X-ray Flux and Fluence of
the Associated Flare
%\\ {\it Solar Physics}
}

\author{C.~\surname{Salas-Matamoros}$^{1,2}$ \sep
        K.-L.~\surname{Klein}$^{1}$     
       }
\runningauthor{Salas-Matamoros et al.}
\runningtitle{CME SXR correlation}

   \institute{$^{1}$ LESIA-UMR 8109 - Observatoire de Paris, CNRS, Univ. Paris 6 \& 7, 92190 Meudon, France \\
                     email: \url{carolina.salas@obspm.fr} email: \url{ludwig.klein@obspm.fr}  \\ 
              $^{2}$ Second affiliation,  Space Research Center, University of Costa Rica
                     email: \url{carolina.salas@cinespa.ucr.ac.cr} \\
             }

\begin{abstract}
Both observation and theory reveal a close relationship between the kinematics of coronal mass ejections (CMEs) and the thermal energy release traced by the related soft X-ray (SXR) emission. The major problem of empirical studies of this relationship is the distortion of the CME speed by the projection effect in the coronagraphic measurements. We present a re-assessment of the statistical relationship between CME velocities and SXR parameters, using the SOHO/LASCO catalog and GOES whole Sun observations during the period 1996 to 2008. 49 events were identified where CMEs originated near the limb, at central meridian distances between 70$^\circ$ and 85$^\circ$, and had a reliably identified SXR burst, the parameters of which - peak flux and fluence - could be determined with some confidence. We find similar correlations between the logarithms of CME speed and of SXR peak flux and fluence as several earlier studies, with correlation coefficients of 0.48 and 0.58, respectively. Correlations are slightly improved over an unrestricted CME sample when only limb events are used. However, a broad scatter persists. We derive the parameters of the CME-SXR relationship and use them to predict ICME arrival times at Earth. We show that the CME speed inferred from SXR fluence measurements
tends to perform better than SoHO/LASCO measurements in the prediction of ICME arrival times near 1 AU. The estimation of the CME speed from SXR observations can therefore make a valuable contribution to space weather predictions.

\end{abstract}
\end{opening}
%-------------------------------------------------

\section{Introduction}
     \label{S-Introduction} 
     
     Coronal mass ejections (CMEs) are expulsions of huge masses of plasma and magnetic field into the heliosphere. When intercepting the Earth, they can trigger geomagnetic storms, {\it i.e}. major disturbances of the terrestrial magnetic field \citep{Gol62,Gon87, Gos93,Zha07,Gop10}. CMEs are often associated with soft X-ray (SXR) bursts \citep{Tan88}, which are routinely observed by the {\it Geosynchronous Operational Environmental Satellites} (GOES) spacecraft. SXR bursts reveal the heating of plasma in a flaring active region. The mechanical energy release to CMEs and the thermal energy release are closely related in many models on the origin of large-scale instabilities in the corona \citep[and references therein]{For06}. Observational studies confirm such a close relationship, when revealing that the acceleration phase of a CME is temporally associated with intense energy release during the rise of the associated SXR burst \citep{Zha01,Zha04,Mar07}. Relationships between the speed or kinetic energy of CMEs on the one hand and the importance of the SXR burst, most often the peak flux, on the other have also been revealed by many statistical studies \citep{Moo03,Bur04,Vrs05,Mar07,Yas09,Bei12}. Occasional negative reports \citep{Agg08} and the broad scatter in the statistical relationship show, however, that the quantitative relationship between CMEs and SXR bursts is not simple.

The interest of clarifying the situation is twofold: on the one hand such statistical relationships show to which extent different manifestations of magnetic energy release in solar eruptions are related. On the other hand empirical relationships between different parameters of solar activity can assist space weather forecasting. This is especially interesting for Earth-directed CMEs whose velocity is not directly measurable by coronagraphs on the Sun-Earth line. Understanding how different quantities describing the output of eruptive solar activity are related is also essential if one wants to use correlation analyses to derive physical relationships with a third quantity, for instance the intensity of solar energetic particle events (see, {\it e.g.}, \citeauthor{tro14}, \citeyear{tro14}). 

A major source of uncertainty in statistical studies involving CME speed is the distortion of the measurement in coronagraphic images by projection effects. \cite{Moo03}, \cite{Bur04} and \cite{Yas09} investigated the above correlations with event samples restricted to CMEs that originated near the solar limb, where projection effects are not expected to affect the CME speed. These authors suggested that the correlations are indeed improved. However, they did not consider the statistical uncertainties of the correlation coefficients. \cite{Yas09} also concluded that the CME speed is more strongly correlated with SXR fluence than with SXR peak flux, but again without addressing the uncertainties in their comparison.

In the present work we re-assess the correlation between CME speed and both SXR peak flux and SXR fluence, restricting ourselves to CMEs near the solar limb. The event selection based on CMEs between 1996 and 2008 from the LASCO CME catalog and the associated GOES SXR bursts is described in Section~\ref{S-Method}. In Section~\ref{S-Res} the results of the statistical analysis are presented, and empirical relationships between CME speed and SXR parameters are derived. The empirical relationships are used in Section \ref{S-Application} in an attempt to predict the arrival times of interplanetary CMEs (ICMEs) at Earth. The results are compared with predictions using CME measurements from SOHO/LASCO and with the observations of ICME arrival near 1~AU.

\section{Methodology and Data Selection}
     \label{S-Method} 
 The data set analyzed in this study consists of parameters of CMEs originating near the solar limb and of the associated SXR bursts. CME parameters (position angles, widths, heights and speeds) are provided in the catalog of CMEs\footnote{\url{http://cdaw.gsfc.nasa.gov/CME_list/}} observed by the {\it Large Angle and Spectrometric Coronagraph} experiment (LASCO; \citeauthor{Bru95}, \citeyear{Bru95}) of the {\it Solar and Heliospheric Observatory} (SOHO), during the period from $1996$ until $2008$. Time histories of SXR flux measured by the GOES satellites in the 0.1-0.8 nm range were retrieved through the database at NASA/GSFC using the IDL routine {\it{goes.pro}} in the {\it SolarSoft} package.

 \subsection{Selection of Limb CMEs}

Limb CMEs were selected in two steps. We first excluded events whose central position Angle (PA, measured counterclockwise from solar north) was within $\pm$ 60$^\circ$ of the projected solar north and south, because such CMEs can only be associated with activity at relatively small central meridian distances. In order to obtain only CMEs with a well-defined direction of propagation, we delimited also the CME width between 60$^\circ$ and 120$^\circ$, especially avoiding halo CMEs. We also excluded CMEs whose speed was $\leq$ 100 km~s$^{-1}$ in order to facilitate the flare association.

For the subsequent correlation studies, we checked the quality of the linear fits to the time-height trajectory and the representativity of the derived CME speed in the LASCO/CME catalog. We found some CMEs whose time-height diagram showed acceleration or deceleration phases in the LASCO field of view (FOV). We included those events where only few points at low altitudes were affected by this acceleration/deceleration, and the linear fit gave a satisfactory estimation of the final speed. In 11 events the acceleration/deceleration was pronounced in the LASCO/C2 FOV. In this case, we used the speed at a distance of 20 solar radii infered from the constant acceleration fit as approximation of the final CME speed.

\begin{figure}    %%%%%%%%%%%%%%%%%% FIGURE 1 
   \centerline{\includegraphics[width=1\textwidth,clip=]{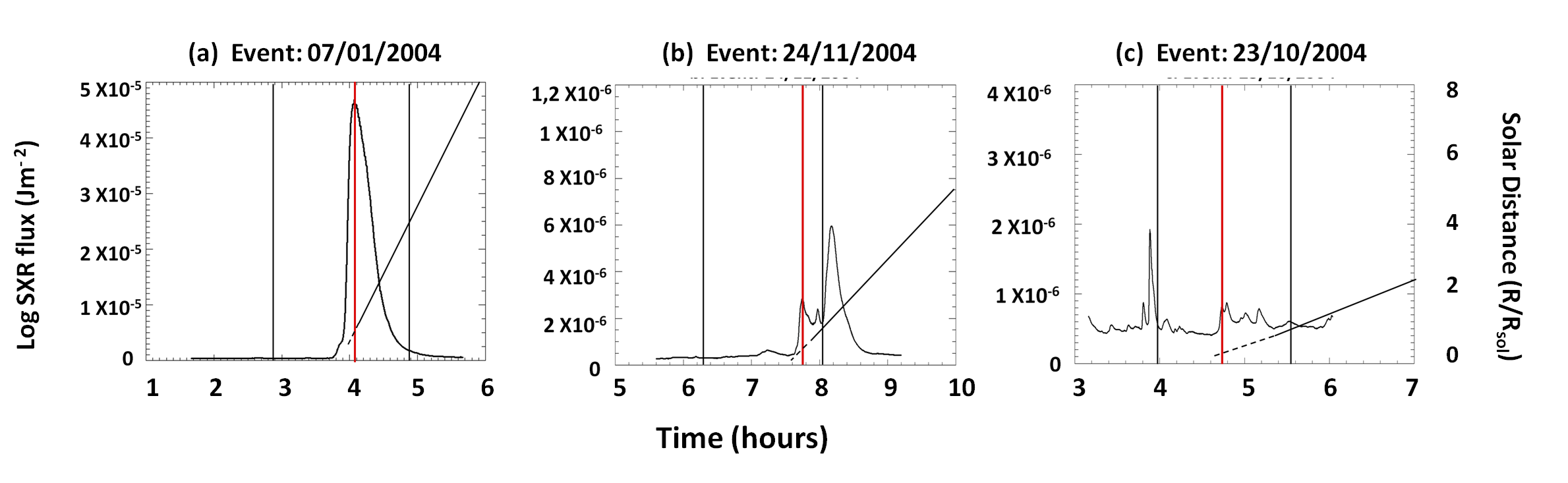}
              }
              \caption{Time profiles of three different SXR bursts:  (a) a well-defined burst with a single peak, (b) a superposition of two different soft X-ray bursts and (c) a burst with a very complex time profile. Vertical black lines delimit the two hour window centered on the time when the extrapolated CME trajectory intersected the solar limb. The vertical red line marks the peak of the SXR burst associated with the CME.} 
   \label{figure1}
   \end{figure}

\subsection{Identification of the Associated Flares}

For the final determination of the origin of CMEs, we identified those associated with flares close to the limb. As a compromise between proximity to the limb and a significant number of events, we focused on flares located, according to {\it Solar Geophysical Data}\footnote{National Geophysical Data Center \url{http://www.ngdc.noaa.gov}}, at central meridian distance between 70$^\circ$ and 85$^\circ$. The events too close to the limb were excluded in order to avoid a partial occultation of the SXR emission. The CME-associated flares were searched in a first step within a fixed time interval with respect to the CME origin. The CME speed (see section 2.1) and the time and heliocentric distance when the CME was first detected were used to extrapolate its trajectory to the limb of the sun. An automated procedure was used to identify SXR bursts that peaked between an hour before and an hour after the instant when the backward extrapolated trajectory intersected the solar limb. This way we identified 77 CMEs associated with flares near the limbs; 44 occurred in the eastern and 33 in the western solar hemisphere.

%-------------- Table of events ----------------------------
% Data from file CMEandFLAREdata.sav
\begin{table}
\footnotesize
\begin{tabular}{l|rrrr|rrrrr}
\hline
\hline
 $N$ & \multicolumn{4}{l}{CME parameters} &  \multicolumn{5}{|l}{SXR parameters} \\
\hline
Date        & $t_0$ & $r(t_0)$  & $V_{\rm CME}$   & $t_{\rm limb}$ &   $t_0$ & $t_p$ &  $F$ & $\Phi_{\rm sp}$ &  Qu \\
%yy mm dd & [hh:mm] & [$R_\odot$] & [km s$^{-1}$] & [hh:mm] & [hh:mm] & [hh:mm] &  [W m$^{-2}$] & [J m$^{-2}$]  & \\
                  &               & [$R_\odot$] & [km s$^{-1}$] &               &               &                &  [Wm$^{-2}$] & [Jm$^{-2}$]  & \\
&  &  &  &    &      & & $(\times 10^5)$ & $(\times 10^4)$  &  \\
(1) & (2) & (3) & (4) & (5) & (6) & (7) & (8) & (9) & (10)  \\
\hline
1996 & & & & & & & & & \\
07 12 & 16:01 & 5.1 & 1085 & 15:17 & 14:59 & 15:32 & 0.49 & 22.80 & 2 \\ 
1997 & & & & & & & & & \\
06 30 & 00:30 & 2.9 & 346 & 23:25  & 23:35 & 23:53 & 0.11 & 4.20 & 1 \\ 
1998 & & & & & & & & & \\
03 13 & 21:30 & 2.7 & 409 & 20:40  & 20:51 & 21:10 & 0.56 & 27.70 & 1 \\ 
04 25 & 15:11 & 2.9 & 349 & 14:09 &  14:02 & 14:37 & 0.36 & 31.30 & 1 \\ 
1999 & & & & & & & & & \\
04 03 & 23:47 & 5.5 & 923 & 22:50 &  22:50 & 23:10 & 4.46 & 137.60 & 1 \\ 
05 08 & 14:50 & 3.8 & 641 & 13:59 &  14:21 & 14:40 & 4.87 & 256.05 & 1 \\ 
05 11 & 22:26 & 4.3 & 735 & 21:34 &  21:25 & 22:05 & 0.40 & 47.80 & 2 \\ 
09 13 & 17:31 & 3.3 & 444 & 16:30 &  17:17 & 17:31 & 0.13 & 5.70 & 2 \\ 
11 08 & 07:26 & 3.5 & 154 & 04:18 &  05:55 & 06:01 & 0.53 & 7.20 & 1 \\ 
2000 & & & & & & & & & \\
06 17 & 03:28 & 4.8 & 857 & 02:36 &  02:19 & 02:37 & 0.38 & 119.30 & 1 \\ 
06 23 & 14:54 & 4.7 & 847 & 14:03 &  14:18 & 14:32 & 3.22 & 120.30 & 1 \\ 
2001 & & & & & & & & & \\
02 03 & 00:30 & 4.0 & 639 & 23:36 &  23:47 & 24:06 & 2.45 & 122.10 & 1 \\ 
04 15 & 14:06 & 4.3 & 1199 & 13:34 &  13:37 & 13:50 & 161.00 & 2708.80 & 1 \\ 
08 10 & 02:06 & 2.5 & 376 & 01:18 &  01:27 & 01:36 & 0.75 & 15.30 & 1 \\ 
10 29 & 08:26 & 2.6 & 617 & 07:56 &  08:00 & 08:13 & 1.08 & 17.96 & 1 \\ 
11 01 & 14:30 & 2.7 & 1053 & 14:11  & 13:50 & 15:01 & 1.26 & 341.20 & 1 \\ 
12 29 & 09:54 & 2.6 & 634 & 09:25 &  09:06 & 09:45 & 9.46 & 316.50 & 2 \\ 
2002 & & & & & & & & & \\
03 13 & 23:54 & 3.6 & 489 & 22:53 &  22:59 & 23:36 & 0.99 & 90.90 & 1 \\ 
04 04 & 05:06 & 2.8 & 468 & 04:22 &  04:12 & 04:40 & 0.87 & 56.00 & 1 \\ 
07 05 & 13:31 & 2.4 & 818 & 13:10 &  12:59 & 13:26 & 3.49 & 124.50 & 1 \\ 
08 03 & 19:31 & 5.2 & 1150 & 18:49 &  19:00 & 19:07 & 11.80 & 137.50 & 1 \\ 
08 16 & 06:06 & 2.5 & 1378 & 05:53 &  05:44 & 06:12 & 2.55 & 193.00 & 1 \\ 
08 22 & 18:26 & 3.0 & 750 & 17:54 &  17:35 & 18:02 & 1.07 & 97.00 & 2 \\ 
08 23 & 13:27 & 2.4 & 321 & 12:38 &  11:41 & 12:00 & 0.88 & 34.70 & 2 \\ 
08 29 & 13:31 & 2.5 & 353 & 12:42 &  12:35 & 12:52 & 3.24 & 75.70 & 1 \\ 
09 08 & 02:06 & 2.5 & 364 & 01:18 &  01:30 & 01:43 & 1.51 & 34.00 & 1 \\ 
10 16 & 04:54 & 2.8 & 250 & 03:30 &  04:05 & 04:23 & 0.21 & 08.70 & 1 \\ 
2003 & & & & & & & & & \\
04 09 & 23:50 & 3.3 & 511 & 22:58 &  23:24 & 23:29 & 2.57 & 21.40 & 1 \\ 
04 25 & 05:50 & 2.9 & 806 & 05:22 &  05:22 & 05:40 & 1.24 & 62.75 & 1 \\ 
10 23 & 20:06 & 2.6 & 1136 & 19:49 &  19:50 & 20:03 & 11.20 & 383.70 & 1 \\ 
10 24 & 02:54 & 2.7 & 1055 & 02:35 &  02:18 & 02:55 & 7.43 & 864.00 & 1 \\ 
11 03 & 10:06 & 2.5 & 1420 & 09:53 &  09:44 & 09:56 & 43.50 & 1404.40 & 1 \\ 
\hline
\end{tabular}
\caption[]{Table of events: date (col.~1), time (2), heliocentric distance (3) of the first detection of the CME in the SoHO/LASCO field of view, speed in the plane of the sky (4), time when the linear backward extrapolation of the time-height trajectory intersected the solar limb (5); times of onset (6), peak (7), peak flux (8), start-to-peak fluence (9) of the SXR bursts, quality flag of the fluence determination (10).}
\label{table0}
\end{table}

\begin{table}
\addtocounter{table}{-1}
\footnotesize
\begin{tabular}{l|rrrr|rrrrr}
\hline
\hline
$N$ & \multicolumn{4}{l}{CME parameters} &  \multicolumn{5}{|l}{SXR parameters} \\
\hline
Date        & $t_0$ & $r(t_0)$  & $V_{\rm CME}$   & $t_{\rm limb}$ &   $t_0$ & $t_p$ &  $F$ & $\Phi_{\rm sp}$ &  Qu \\
%yy mm dd & [hh:mm] & [$R_\odot$] & [km s$^{-1}$] & [hh:mm] & [hh:mm] & [hh:mm] &  [W m$^{-2}$] & [J m$^{-2}$]  & \\
                  &               & [$R_\odot$] & [km s$^{-1}$] &               &               &                &  [Wm$^{-2}$] & [Jm$^{-2}$]  & \\
&  &  &  &    &      & & $(\times 10^5)$ & $(\times 10^4)$  &  \\
(1) & (2) & (3) & (4) & (5) & (6) & (7) & (8) & (9) & (10)  \\
\hline
2004 & & & & & & & & & \\
01 07 & 04:06 & 3.0 & 1581 & 03:51 &  03:42 & 04:03 & 4.65 & 170.00 & 1 \\ 
01 07 & 10:30 & 3.5 & 1822 & 10:14 &  10:15 & 10:26 & 8.46 & 219.40 & 1 \\ 
05 07 & 10:50 & 3.2 & 469 & 09:55 &  09:53 & 10:19 & 0.07 & 6.01 & 2 \\ 
05 17 & 05:26 & 2.8 & 383 & 04:31 &  04:11 & 04:17 & 0.79 & 5.60 & 1 \\ 
06 16 & 04:36 & 2.7 & 603 & 04:03 & 03:59 & 04:30 & 0.27 & 14.50 & 1 \\ 
08 24 & 13:54 & 3.3 & 817 & 13:22 &  13:30 & 13:49 & 0.06 & 4.10 & 1 \\ 
08 31 & 05:54 & 2.4 & 311 & 05:00 &  05:19 & 05:38 & 1.50 & 47.10 & 1 \\ 
11 03 & 02:06 & 2.4 & 379 & 01:22 &  00:50 & 01:33 & 3.00 & 82.50 & 1 \\ 
11 24 & 22:06 & 2.5 & 262 & 21:01 &  21:29 & 21:45 & 0.96 & 40.90 & 1 \\ 
2005 & & & & & & & & & \\
04 17 & 21:26 & 2.9 & 721 & 20:54 &  20:41 & 21:07 & 0.48 & 26.90 & 2 \\ 
05 06 & 03:30 & 4.0 & 1120 & 02:59 &  03:06 & 03:13 & 0.88 & 19.20 & 1 \\ 
05 06 & 11:54 & 5.8 & 1144 & 11:05 &  11:12 & 11:28 & 1.30 & 34.60 & 1 \\ 
08 25 & 04:54 & 4.2 & 1327 & 04:26 &  04:33 & 04:40 & 6.63 & 93.10 & 1 \\ 
09 04 & 14:48 & 2.5 & 1179 & 14:33 &  13:59 & 15:07 & 0.22 & 39.50 & 1 \\ 
2006 & & & & & & & & & \\
04 29 & 16:54 & 2.5 & 491 & 16:18 &  16:10 & 16:30 & 0.23 & 8.60 & 1 \\ 
04 30 & 02:06 & 2.5 & 428 & 01:26 &  01:32 & 01:57 & 0.53 & 35.96 & 2 \\ 
2008 & & & & & & & & & \\
03 25 & 19:31 & 5.8 & 1103 & 18:40 &  18:39 & 18:56 & 1.72 & 74.20 & 1 \\ 
\hline
\end{tabular}
\caption[]{Table of events (cont'd).}
\end{table}
% ---------------End table------------------------- 

The time profile of each SXR burst of this sample was studied in detail to identify cases when the CME-flare association found by the automated search was ambiguous. 
We discarded weak bursts, because they would not allow us to obtain reliable values of the fluence.

Three different types of time profiles were identified (see Figure \ref{figure1}): (a) a well-defined peak, (b) a burst with more than one peak, which may mean a superposition of different bursts, and (c) a very complex profile where no burst could be unambiguously associated with the CME. The events in the latter category were discarded.  For the cases with several peaks, we verified the coordinates of the flare related to each peak in the time profile directly through the analysis of image sequences from SOHO/EIT 19.5 nm \citep{Del95} or {\it Yohkoh}/SXT \citep{Tne91}. The events where images revealed flares in active regions within $\pm$ 70$^\circ$ from the central meridian or at the opposite limb of the CME were eliminated, as well as events where several peaks were associated to the same active region without possibility to distinguish if one or several were actually associated with the CME. We also discarded cases when the CME reported in the catalog was not clearly recognizable in the LASCO daily movies. We eventually obtained a list of 49 events for which the correlation between CME speed and SXR peak flux and fluence could be studied. They are listed in Table~\ref{table0}. The fluence calculation will be discussed in Section~\ref{S-Res}. 

The CME speeds in the sample range from 154 to 1822~km~s$^{-1}$, with a median of 639~km~s$^{-1}$, the SXR peak fluxes from $6 \cdot 10^{-7}$ to $1.6 \cdot 10^{-3}$~W~m$^{-2}$, {\it i.e}. from GOES classes B6 to X16.

\section{Correlation between CME Speed and SXR Peak Flux and Fluence}
     \label{S-Res} 
    \begin{figure}    %%%%%%%%%%%%%%%%%% FIGURE 2 
   \centerline{\includegraphics[width=0.9\textwidth,clip=]{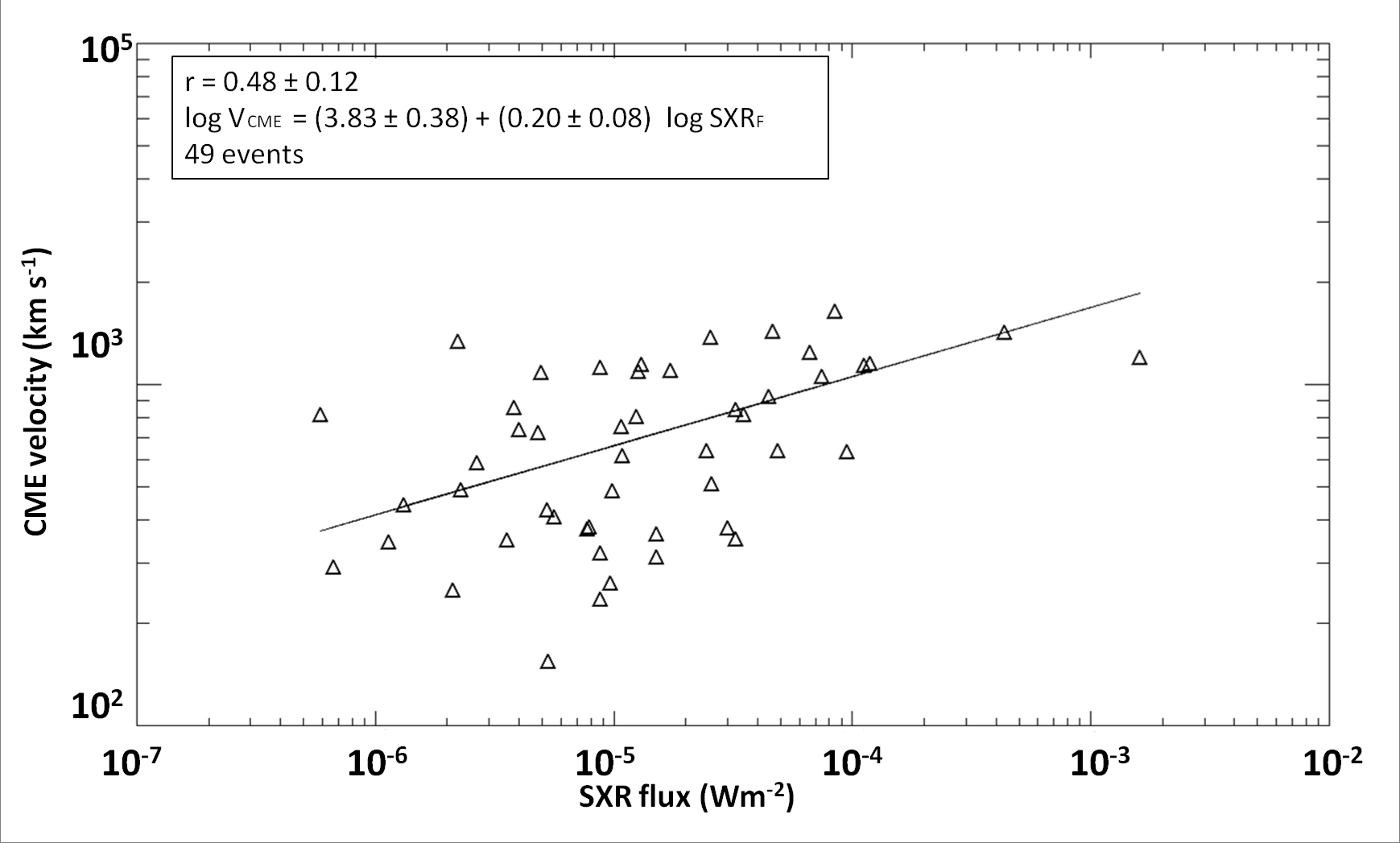}
              }
              \caption{The logarithmic plot of the speed of CMEs near the solar limb during the period 1996-2008 versus the SXR peak $F$ of the associated flares. The straight line is the result of a least absolute deviation fit. The insert shows the correlation coefficient, the parameters of the straight line, and the number of events.
              }
   \label{figure2}
   \end{figure}
   
\begin{figure}    %%%%%%%%%%%%%%%%%% FIGURE 3 
   \centerline{\includegraphics[width=0.9\textwidth,clip=]{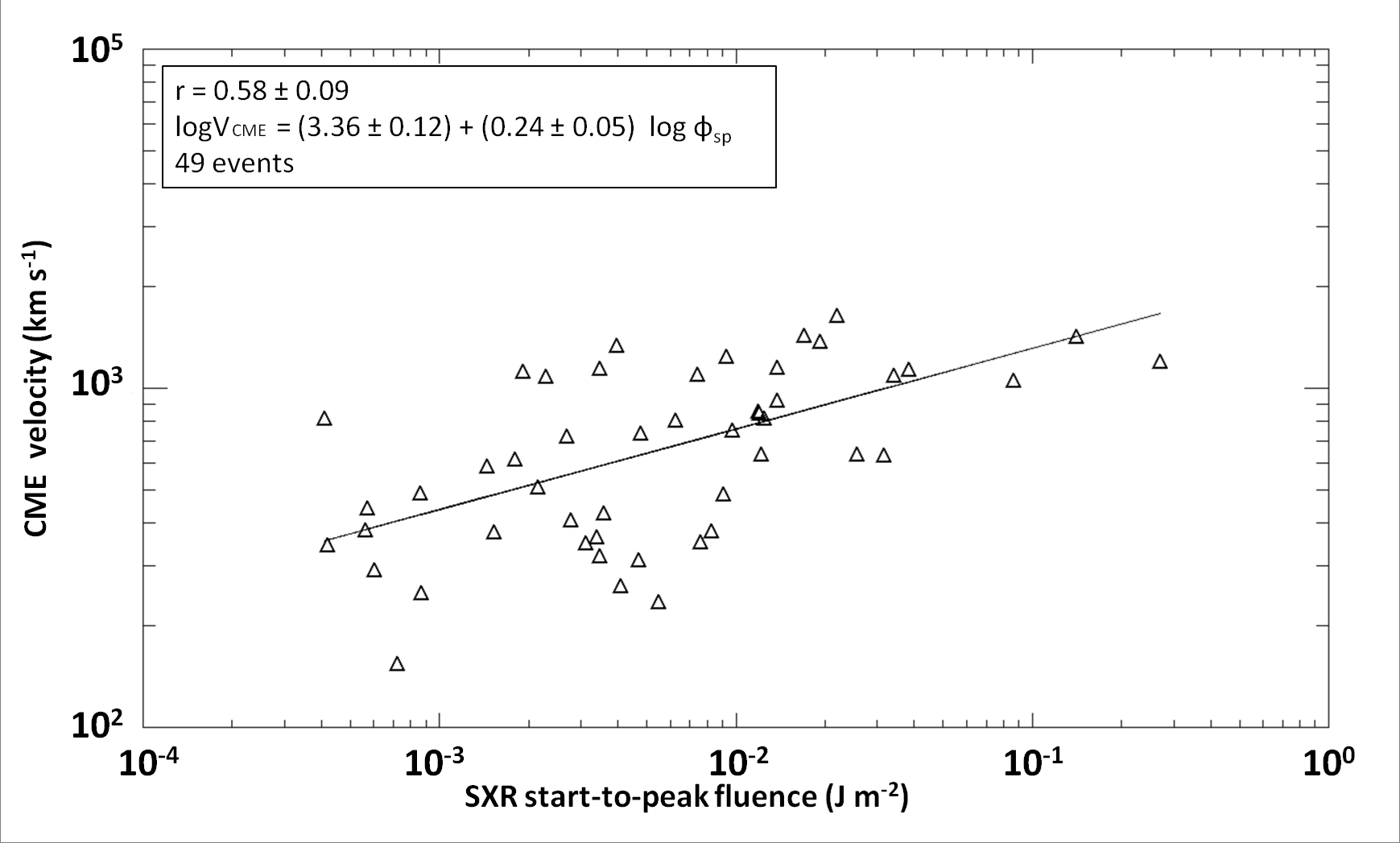}
              }
              \caption{The logarithmic plot of the speed of CMEs near the solar limb during the period 1996-2008 versus the SXR start-to-peak fluence $\Phi_{\rm sp}$ of the associated flares. See caption of Figure~\ref{figure2}.
              }
   \label{figure3}
   \end{figure}
 
Based on the new filtered list of 49 events (25 at the eastern and 24 at the western limb), we related the speeds of the CMEs with parameters of the associated SXR bursts as observed by GOES in the 0.1-0.8~nm channel. Figure ~\ref{figure2} displays the scatter plot of the CME speed {\it vs} the SXR peak flux on a double-logarithmic scale. We found a positive correlation of $r= 0.48 \pm 0.12$ between the logarithms of the CME speed and of the SXR peak flux. 
Here and in the following the errors were calculated using a bootstrap method \citep[ch. 6.6]{Wal03}: the correlation coefficient was calculated repeatedly for a randomly selected sample of 49 out of the 49 observed data pairs, and the mean and standard deviation are quoted as the correlation coefficient and its statistical uncertainty.

Besides the peak flux we also considered the fluence. Two types of fluence were calculated in the 0.1-0.8 nm band for these events, namely start-to-peak fluence and total fluence. The background was determined as the average flux in a suitable time interval before the SXR burst. The start-to-peak fluence was calculated by integrating the background-subtracted flux from the start of the SXR burst until its maximum, including possible small previous peaks that we considered as precursors. The existence of such previous peaks, and problems with background determination introduce uncertainties inside the fluence calculation. The quality flag in col.  10 of Table \ref{table0} is an assessment based on visible inspection. Qu=1 means that the fluence is reliable, Qu=2 labels less certain cases.
 
The total fluence is more difficult to calculate, because the end of the SXR burst is generally not well defined, and new events may be superposed on the decay of the burst of interest. \cite{Kal89} defined the end of the burst as the time when the X-ray flux returns to the GOES C2 level, while \cite{Yas09} used the time when the soft X-ray flux decays to half of the peak value. We fitted the decay from the main peak by an exponential and calculated the fluence analytically until infinity. This avoids contamination by new SXR bursts during the decay phase as well as an arbitrary definition of the end time.

We obtained the same correlation between the CME speed and the SXR start-to-peak fluence and total fluence, $r= 0.58 \pm 0.09$. The probability to obtain this or a higher correlation coefficient from an unrelated sample is $1.3\cdot 10^{-5}$. The result is similar to those of \cite{Moo02} and \cite{Yas09} who found correlations of 0.47 and 0.56, respectively.

The relationship between the logarithms of the CME speed $V_{\rm CME}$ and of the peak flux $F$ and fluence (start-to-peak fluence $\phi_{\rm sp}$ and total fluence $\phi_{\rm p}$) of the associated SXR burst were inferred using linear fits minimizing least squares deviation and least absolute deviation. Differences  between the resulting velocities amounted up to some tens of km~s$^{-1}$ in extreme cases. Although these differences are small compared with the overall statistical uncertainty, we use in the following the result from the least absolute deviation fit, which is less sensitive to outliers. This leads to the following empirical relationships:
 
%\begin_inset Formula for flux
\begin{equation}
\log V_{\rm CME}=(0.20\pm0.08)\log F+(3.83\pm0.38) \; ,
\label{relationshipflux}
\end{equation}
%\end_inset

%\begin_inset Formula for fluence until peak
\begin{equation}
\log V_{\rm CME}=(0.24\pm0.05)\log \phi_{\rm sp}+(3.36\pm0.12) \; ,
\label{relationshipfluence}
\end{equation}
%\end_inset

%\begin_inset Formula for fluence with analitical and numerical calculation
\begin{equation}
\log V_{\rm CME}=(0.22\pm0.05)\log \phi_{\rm p}+(3.21\pm0.10) \; .
\label{relationshipAn_fluence}
\end{equation}
%\end_inset
%The errors were calculated using a bootstrap method \citep[ch. 6.6]{Wal03}. 

\noindent These results are independent of whether we use all events or only those with quality flag 1.

Our analysis is simplified in several respects. We used a standard minimization technique that is in principle justified only when all the measurement uncertainties are in the dependent variable, here the CME speed, whereas the independent variable is supposed to be exactly known. This is of course not the case, and we would have to apply a more general technique, such as total least squares minimization. We checked this and found no significant difference with the results of the standard fits above. 

A second problem is a bias in our statistics, due to our rejection of CMEs that were accompanied by weak or undetected SXR bursts. The fitted straight line would be expected to have a steeper slope if these events, which are located in the lower left corner of Figures~\ref{figure2} and ~\ref{figure3}, had been considered. We found indeed that the straight line steepened when we gradually extended the minimum fluence considered from $10^{-2}$~J~m$^{-2}$ to the lowest value detected, and it would likely steepen more than indicated by Equation~\ref{relationshipfluence} if weak SXR bursts were not hidden in the background. The above relationships may hence overestimate the speeds of CMEs associated with weak SXR bursts and underestimate those of CMEs with intense SXR emission.

\section{Application of the CME-SXR Relationship to ICME Propagation}
     \label{S-Application} 
        \begin{sidewaystable}
\tiny

%\begin{table}
%\footnotesize 

\begin{tabular}{c|cc|ccc|cccc}

\hline %linea horizontal
$N$ &
\multicolumn{2}{c}{ICME onset} &
\multicolumn{3}{c}{LASCO observations} &
\multicolumn{4}{c}{SXR bursts} \\

&
{\it Wind} &
ACE &
Date/ &
Speed &
Predicted &
Start &
$\Phi_{\rm sp}$ &
Speed &
Predicted \\

&
ICME &
ICME &
Time &
[km s$^{-1}$] &
arrival &
&
[J m$^{-2}$] &
[km s$^{-1}$] &
arrival \\
\hline %linea horizontal

&
1997 &
&
&
&
&
&
&
&
 \\

1 &
11 Apr 09:00 &
- &
7 Apr 14:27 &
830 &
10 Apr 18:00   &
14:08 &
0.0036 &
592 &
11 Apr 09:08 (10 Apr 12:42 - 11 Apr 20:50)\\

2 &
15 May 10:30 &
- &
12 May 06:30 &
306 &
16 May 19:30 &
04:55 &
0.0004 &
355 &
16 May 14:55 (16 May 01:08 - 16 May 23:13)\\ 

&
1998 &
&
&
&
&
&
&
&
 \\

3 &
24 Jun 18:00 &
24 Jun 17:00 &
21 Jun 05:35 &
307 &
25 Jun 18:25 &
05:18 &
0.0043 &
619 &
24 Jun 22:18 (24 Jun 01:00 - 25 Jun 10:43)\\ 

&
1999 &
&
&
&
&
&
&
&
 \\

4 &
16 Apr 19:00 &
16 Apr 18:00 &
13 Apr 03:30 &
282 &
17 Apr 18:00 &
02:14 &
0.0002 &
309 &
17 Apr 15:15 (17 Apr 03:04 - 17 Apr 22:58)\\ 

&
2000 &
&
&
&
&
&
&
&
 \\

5$^{(*)}$ &
- &
8 Jun 12:30 &
6 Jun 15:54 &
1098 &
8 Jun 22:05 &
15:26 &
0.2002 &
1557 &
8 Jun 02:26 (7 Jun 15:42 - 8 Jun 17:19)\\

6 &
24 Jun 07:00 (f)&
24 Jun 03:00 &
20 Jun 09:10 &
471 &
24 Jun 11:40 &
08:27 &
0.0020 &
516 &
24 Jun 08:00 (23 Jun 15:00 - 24 Jun 18:57)\\

7 &
28 Jul 14:00 &
28 Jul 13:00 &
25 Jul 03:30 &
532 &
29 Jul 02:05 &
02:50 &
0.0077 &
712  &
28 Jul 13:50 (27 Jul 13:23 - 29 Jul 03:39)\\

8 &
31 Jul 23:00 &
31 Jul 22:00 &
28 Jul 18:30 &
832 &
31 Jul 21:50 &
17:16 &
0.0003 &
314 &
2 Aug 04:30 (1 Aug 15:29 - 2 Aug 12:38)\\

&
2001 &
&
&
&
&
&
&
&
 \\
 
9 &
21 Oct 20:00 &
- &
19 Oct 16:50 &
901 &
22 Oct 13:50 &
16:15 &
0.0668 &
1197 &
21 Oct 16:45 (21 Oct 00:56 - 22 Oct 14:54)\\

10 &
6 Nov 12:00 &
- &
4 Nov 16:35 &
1810 &
5 Nov 21:25 &
16:03 &
0.0460 &
1094 &
6 Nov 22:30 (6 Nov 04:14 - 7 Nov 21:35)\\

&
2002 &
&
&
&
&
&
&
&
 \\

11  &
8 Sep 04:00 &
8 Sep 04:00 &
5 Sep 16:54 &
1748 &
6 Sep 22:54 &
16:19 &
0.0073 &
704 &
9 Sep 04:00 (8 Sep 03:41 - 9 Sep 17:35)\\

&
2003 &
&
&
&
&
&
&
&
 \\

12 &
28 Oct 02:30 &
- &
26 Oct 17:54 &
1537 &
28 Oct 05:10 &
17:15 &
0.2230 &
1598 &
28 Oct 02:45 (27 Oct 16:49 - 28 Oct 17:25)\\

&
2004 &
&
&
&
&
&
&
&
 \\
13 &
22 Jul 18:00 &
- &
20 Jul 13:31 &
710 &
24 Jul 00:45 &
12:17 &
0.0190 &
885 &
23 Jul 10:45 (22 Jul 10:19 - 24 Jul 04:26)\\

&
2005 &
&
&
&
&
&
&
&
 \\

14 &
21 Jan 19:00 &
21 Jan 19:00 &
20 Jan 06:54 &
882 &
23 Jan 05:45 &
06:32 &
0.5306 &
1968 &
21 Jan 08:30 (21 Jan 01:10 - 21 Jan 19:04)\\

15 &
15 May 06:00 &
15 May 06:00 &
13 May 17:12 &
1689 &
15 May 00:30 &
16:28 &
0.0615 &
1173 & 
15 May 18:10 (15 May 01:54 - 16 May 16:50)\\

&
2006 &
&
&
&
&
&
&
&
 \\

16 &
14 Dec 22:00 &
14 Dec 22:00 &
13 Dec 02:54 &
1774 &
14 Dec 08:25 &
02:18 &
0.2243 &
1600 &
14 Dec 11:50 (14 Dec 01:50 - 15 Dec 02:23)\\

17 &
17 Dec 00:00 &
17 Dec 00:00 &
14 Dec 22:30 &
1042 &
17 Dec 08:30 &
22:00 &
0.0451 &
1089 &
17 Dec 04:50 (16 Dec 10:23 - 18 Dec 03:48)\\

&
2010 &
&
&
&
&
&
&
&
 \\

18 &
5 Apr 12:30 &
5 Apr 12:00 &
3 Apr 10:30 &
668 &
7 Apr 00:30 &
09:53 &
0.0012 &
453 &
7 Apr 13:20 (6 Apr 21:17 - 7 Apr 23:14)\\

&
2011 &
&
&
&
&
&
&
&
 \\

19 &
18 Feb 06:00 &
18 Feb 05:30 &
15 Feb 02:24 &
669 &
18 Feb 16:24 &
01:56 &
0.0570 &
1152 &
17 Feb 04:56 (16 Feb 12:04 - 18 Feb 03:52)\\

20 &
5 Aug 04:00 (f)&
5 Aug 03:00 &
2 Aug 06:39 &
712 &
5 Aug 17:50 &
05:48 &
0.0118 &
790 &
5 Aug 12:05 (4 Aug 10:08 - 6 Aug 02:46)\\

21 &
10 Sep 03:30 (f)&
10 Sep 05:00 &
6 Sep 22:30 &
575 &
10 Sep 18:30 &
22:02 &
0.0241 &
937 &
9 Sep 16:02 (8 Sep 17:11 - 10 Sep 11:33)\\

&
2012 &
&
&
&
&
&
&
&
 \\

22 &
23 Jan 00:30 &
22 Jan 23:30 &
19 Jan 15:09 &
1120 &
21 Jan 20:00 &
14:28 &
0.1002 &
1319 &
21 Jan 09:20 (20 Jan 19:45 - 22 Jan 04:52)\\

23 &
9 Mar 04:30 &
- &
7 Mar 00:24 &
2684 &
7 Mar 18:24 &
00:00 &
0.2770 &
1683 &
8 Mar 07:30 (7 Mar 22:13 - 8 Mar 20:54)\\

24 &
12 Mar 22:00 (f)&
- &
10 Mar 17:54 &
491 &
14 Mar 19:05 &
17:15 &
0.0772 &
1239 &
12 Mar 15:45 (12 Mar 00:41 - 13 Mar 12:55)\\

25 &
16 Jun 23:00 &
16 Jun 22:30 &
14 Jun 14:09 &
987 &
17 Jun 04:10 &
12:50 &
0.0431 &
1077 &
16 Jun 20:20 (16 Jun 01:40 - 17 Jun 19:14)\\

26 &
15 Jul 07:00 &
15 Jul 06:30 &
12 Jul 16:54 &
885 &
15 Jul 15:25 &
15:42 &
0.1936 &
1545 &
14 Jul 02:40 (13 Jul 16:11 - 14 Jul 18:08)\\

\hline %linea horizontal
\end{tabular}

\caption{Comparison of travel time of CMEs based on {\it Wind} and ACE measurements and based on inferred speeds using the empirical interplanetary propagation model. An asterisk (*) in col. 1 indicates that the ICME arrival is uncertain. Suffix(f) in col. 2 indicates that probably only the flank of the ICME passed over the spacecraft.}

\label{table4}

%\end{table}

\end{sidewaystable}

In this section we test the relationship between CME speed and SXR fluence by applying it to the prediction of ICME arrival times at Earth.

The arrival of an ICME at Earth is one of the rare issues of space weather where the Sun leaves a substantial warning time. Yet the prediction of the arrival time is difficult: on the one hand the speeds of Earth-directed CMEs cannot be directly measured by a coronagraph on the Earth-Sun line. On the other hand the CME is not a rigid object, but changes during propagation in the interplanetary medium, where CMEs expand, change shape due to compression and reconnection, and are accelerated or decelerated. The relevant processes are reviewed, {\it e.g.}, in \cite{For06} and \cite{Dem10}. Detailed analyses using heliospheric imaging from STEREO were reported by \cite{Col13} and \cite{Mos14}; see also the review of \cite{Rou11}.

Many attempts were undertaken in the literature to derive simple methods to forecast ICME arrival times at the Earth using CME observations at the Sun. These models must take account of the acceleration or deceleration of CMEs in the interplanetary medium \citep{Gop01, Sch05,Vrs10}.

\cite{Gop01} proposed a simple analytical treatment of the interplanetary propagation, based on an empirical relationship between the acceleration, assumed constant out to a limiting heliocentric distance, and the radial front speed of the CME in the corona. We applied the empirical relationship from their Equation~4, which can be formulated as

\begin{displaymath}
a \left[ {\rm m \; s^{-2}} \right] =  -0.0054 \left(V_{\rm CME} - 406 \; \left[{\rm km \; s ^{-1}}\right] \right) \; .
\end{displaymath}

We suppose that the acceleration ceases either when the ICME attains the speed of 406 ${\rm km \; s ^{-1}}$ or at the latest when it is at heliocentric distance 0.76 AU. This differs slightly from \cite{Gop01} who considered that the acceleration or deceleration always continued out to 0.76 AU. For CME speeds below 800 ${\rm km \; s ^{-1}}$ the travel times derived from the two methods differ by a few hours. We will refer to the model as empirical interplanetary propagation model in the following. We estimated the CME speed in the corona in two different ways: firstly, using the speed measurements from LASCO and secondly, using Equation~\ref{relationshipfluence} to infer the CME speed from the SXR fluence.

\begin{figure}    %%%%%%%%%%%%%%%%%% FIGURE 4  
  \centerline{\includegraphics[width=0.9\textwidth,clip=]{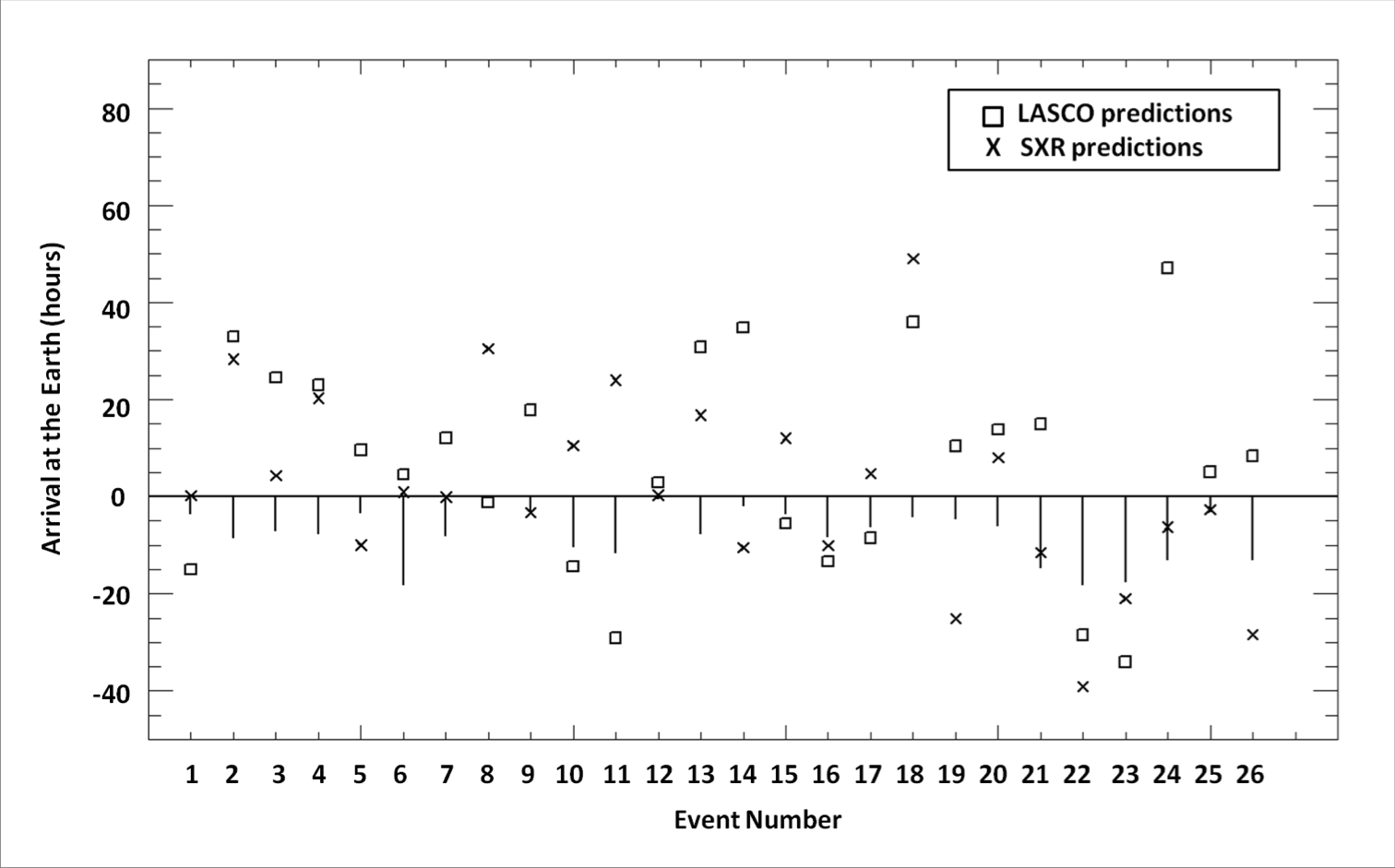}
              }
              \caption{Representation of the predictions of arrival at the Earth with reference to the observed ICME arrival (0 on the ordinate). The vertical lines indicate the time interval between the shock arrival and the ICME arrival at the {\it Wind} spacecraft.}
   \label{figure4}
  
   \end{figure}

\subsection{Results}

We compared the predicted arrival time with observations at {\it Wind} and  ACE for a ~list of selected ICMEs with well-observed arrival times at the spacecraft. We used 26 ICMEs listed by \cite{Gop01}, in the online catalog of Richardson and Cane\footnote{\url{http://www.srl.caltech.edu/ACE/ASC/DATA/level3/icmetable2.htm}}, and by \cite{Mos14}.

The predicted arrival times were compared with the observed arrival of both the shock and the driver. The driver is considered to be the ICME. While the shock arrival at the spacecraft was usually well determined by a sudden increase of the temperature, density, and magnetic field intensity, the arrival of the ICME was often ambiguous and may depend on the parameter used to identify it. We employed one or a combination of the following: the start of a magnetic field enhancement, of a depression of proton temperature or the proton plasma beta, of a gradually decreasing high solar wind speed or of magnetic field rotation. 

The 26 CME/ICME pairs displayed in Table \ref{table4} are those for which we could (i) confirm the onset time identified in the published lists to within one or two hours, (ii) clearly associate a SXR burst with the CME. ICMEs where such bursts could not be identified were discarded ({\it e.g.}, ICMEs on 10 January and 10 February 1997).  

The first column of Table \ref{table4} shows the event number followed by the times of ICME arrival identified from {\it Wind} and ACE measurements. In all cases but event 5 we used the ICME arrivals from {\it Wind}. In four cases (6, 20, 21, and 24) only the flank of the ICME passed over the spacecraft making the determination of the arrival time uncertain. These events are identified with a label ``f" in Table~\ref{table4} after the date. The next three columns summarize the CME data from the LASCO catalog and the predicted arrival time of the ICME at the Earth using the LASCO speed as input to the empirical interplanetary propagation model, and taking as reference the heliocentric distance and the time of the first detection of the CME by LASCO as given in the catalog. The last columns give the start time and start-to-peak fluence of the related SXR bursts, the CME speed inferred from the fluence and the arrival time of the ICMEs as calculated by the propagation model. The reference is the start time of the burst. Values whithin parentheses give the uncertainty interval of the expected ICME arrival due to the uncertainty of the coefficients of Equation \ref{relationshipfluence}.

\begin{figure}    %%%%%%%%%%%%%%%%%% FIGURE 5  
  \centerline{\includegraphics[width=1\textwidth,clip=]{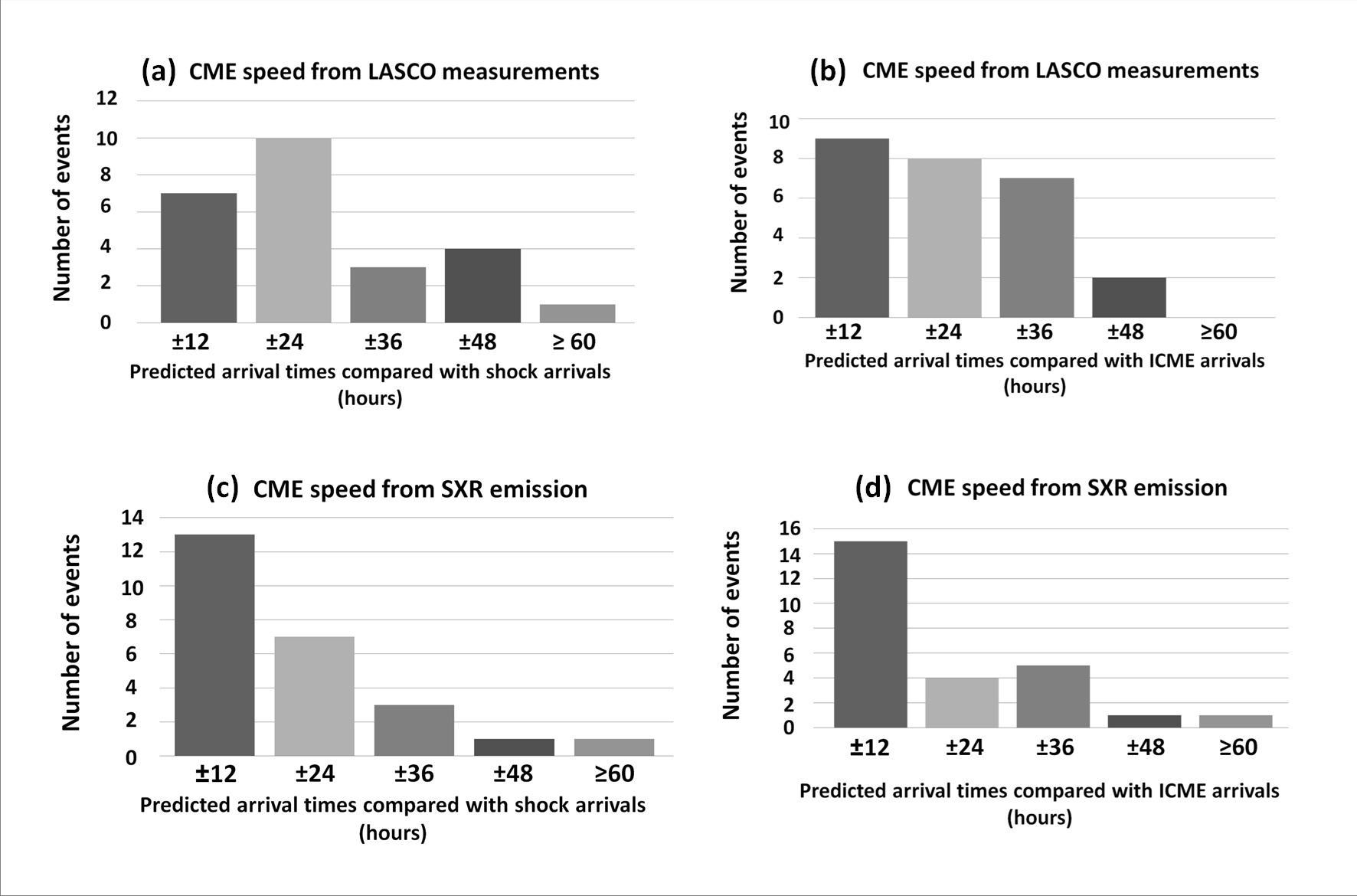}
              }
              \caption{Comparison of predicted ICME arrival at 1 AU with the observed onset of the shock and ICME at the {\it Wind} spacecraft. The predictions are compared with the observed arrival of the shock in histograms (a) and (c), and of the ICME  in histograms (b) and (d).  Histograms in the top row show the predictions using LASCO measurements, those in the bottom row predictions using the propagation speed inferred from SXR fluence.}
   \label{figure5}
  
   \end{figure}

A graphical comparison between the predicted and observed arrival times is given in Figure~\ref{figure4}. The reference zero of the vertical axis is the time when the ICME, {\it i.e.} the driver, reached the {\it Wind} spacecraft. The vertical bars indicate the time interval between the arrival of the shock and the driver, that is, the size of the sheath region. The arrival time predicted using the LASCO CME speed is represented by an open square, the prediction using the propagation speed inferred from Equation~\ref{relationshipfluence} by a cross.

On average we observe that the arrival times predicted using the SXR parameters are closer to the observed arrival times than those predicted using LASCO measurements. Figure~\ref{figure5} gives another comparison between the two predictions of the ICME arrival time and the observations in panels (b) and (d). The comparison with the observed shock arrival time is given in panels (a) and (c). Predictions using the LASCO observations  are shown in the top row, those based on the SXR fluence in the bottom row. The events are grouped into 12~h intervals with respect to the arrival of the ICME shock  (a,c) and the driver (b,d). The first bar hence gives the number of events where the absolute value of the delay between the predicted and observed arrival is greater or equal to 0 and less than 12 h, {\it etc}. The figure confirms the impression from Figure~\ref{figure4} that ICME travel times estimated from the SXR fluence tend to cluster more closely around the arrival times of both the shock and the driver than the travel times inferred from the LASCO measurements. 
In 15/26 events the SXR-inferred CME speed leads to an ICME arrival prediction within $\pm 12$ h of the observed time. Only 9/26 cases where coronagraphic observations are used achieve this. The median error of the prediction from SXR fluence is 11.5 h and from LASCO predictions, 14.5 h. Caution is of course necessary because of the small event sample.

\subsection{Assessment of Failed Predictions}

In 8/26 events the observed arrival time of the ICME is outside the range of uncertainty of the SXR fluence prediction (events 2, 4, 8, 18, 19, 22, 23 and 26). This set includes the six events in the three highest bins of Figure 5(d), and two other events where the ICME arrival prediction was wrong by more than 20 h. Five of these events are also poorly predicted when the CME speed from LASCO is used, while in the three others (8, 19, 26) speeds from LASCO observations lead to a better estimate of the ICME arrival than the estimation based on the SXR fluence. 

In some of the events we obtained an over estimation or under estimation of the speed that affected the predictions of ICME arrival. In the events 2, 4, and 8 we found low speeds of 355, 309, and 314 km~s$^{-1}$, respectively, with a corresponding delay of the ICME arrival of 29, 20, and 29 h, respectively. The LASCO measurements were similarly slow for the events 2 and 4, but not for the event 8, where the observed speed was 832 km~s$^{-1}$ providing a prediction in advance by only 1 hour of the observed ICME arrival. 

In the remaining events we can use published observation from the {\it Solar Terrestial Relations Observatory} (STEREO) for a more detailed assessment of the failed predictions. 
In the case of the event 18 on 3 April 2010, the SXR prediction is late by 48 h, while LASCO is late by 36 h. The studied CME is moderately fast, with a higher speed observed by LASCO (668 km~s$^{-1}$) than inferred from the SXR fluence (456 km~s$^{-1}$). This event was observed by STEREO B at the limb with a speed of 833 km~s$^{-1}$ \citep{Woo11}. When this speed is used in the ICME propagation model, an interplanetary travel time of about 51 h and an ICME arrival at 1 AU near 12 UT on 5 April is predicted, which is in excellent agreement with the observations. So the failed ICME prediction based on the speeds from LASCO and from the SXR fluence is most likely due to the erroneous estimates of the Earth-directed CME speed. 

The SXR prediction of the ICME arrival for the event 19 on 15 February 2011 is early by 25 h, while the prediction from LASCO measurements is late by 10 h. The CME speed inferred from the SXR fluence is higher (1152 km~s$^{-1}$) than from LASCO observations (669 km~s$^{-1}$). An intermediate CME speed of 945 km~s$^{-1}$ was measured by STEREO A, where the event occurred near the limb \citep{Schr11}. The travel time to 1 AU is about 65 h, predicting the arrival of the ICME on 17 February near 19 UT, that is about 6 h too early. On the other hand, the three-dimensional (3-D) modeling by \cite{Tem14} and \cite{Mis14} gave initial CME speeds of about 1000-1100 km~s$^{-1}$, in good agreement with the speed inferred from SXR fluence. \cite{Mis14} reported a pronounced deceleration from 1100 km~s$^{-1}$ at 6 $R_\odot$ to 580 km~s$^{-1}$ at 11 $R_\odot$. This suggests that in this case the CME speed inferred from the SXR fluence was an adequate estimate, but the interplanetary transport was complex, probably due to the interaction with previous CMEs \citep{Tem14,Mis14}.

In the event 22 on 19 January 2012, the SXR fluence-based prediction is early by 39 h, LASCO by 28 h. The CME is fast, with a lower speed estimate from LASCO observations (1120 km~s$^{-1}$) than from the SXR fluence (1319 km~s$^{-1}$). A CME speed of 1335 km~s$^{-1}$ was inferred from 3-D modeling \citep{Mos14}, confirming our estimation from the SXR fluence. The failure of our arrival predictions is hence not likely to be due to erroneous estimate of the CME speed in the corona. The detailed analysis of the CME and its interplanetary propagation \citep{Liu13} reveals a rapid deceleration of the CME down to 700-800 km~s$^{-1}$ within 35 $R_\odot$ from the Sun, and a subsequent propagation at roughly constant speed. The simple propagation model applied in the present study predicts such speeds only at the imposed terminal distance of 0.76 AU, and therefore underestimates the interplanetary travel time.

The CME in the event 23 on 7 March 2012 is very fast, with a higher speed observed by LASCO (2684 km~s$^{-1}$) than inferred from the SXR fluence (1683 km~s$^{-1}$). A similarly high speed as in the LASCO measurement (2585 km~s$^{-1}$) was found in the 3-D modeling \citep{Mos14}. But this CME has a complex propagation into the interplanetary medium \citep{Rol14} The speed inferred from SXR fluence underestimates the CME speed. 
On the other hand, a deceleration of this ICME in the interplanetary space was observed by \cite{Liu13}, \cite{Dav13} and \cite{Rol14}. The analyses of \cite{Liu13} and \cite{Rol14} suggest that the deceleration was enhanced by the interaction of the fast CME with previous ones. The interplanetary propagation cannot be described by a simple empirical propagation model in this case.

Finally, for the event 26 (12 July 2012), the prediction of arrival time of the ICME based on the SXR fluence is early by 28 h, and that based on LASCO observations by 8 h. The CME speed inferred from SXR fluence is high (1545 km~s$^{-1}$), while the LASCO measurement is 885 km~s$^{-1}$. From the analysis of STEREO observations with a drag model of interplanetary transport, \cite{Hes14} derived an initial speed of 1316 km~s$^{-1}$. This speed would predict a travel time of about 43 h and an ICME arrival near 12 UT on 14 July, which is well in advance of the observed arrival. The issue is hence rather one of the interplanetary propagation of the CME than of the speed determination from the SXR fluence, which is closer to the result of the STEREO observations than the speed from LASCO.

\section{Summary and Discusion}
     \label{S-Discusion}      
  \subsection{Summary of Observational Results}

The re-assessment conducted in the present work of the correlation between the speed of a CME near the limb and the parameters of the associated SXR burst, provided such a burst can be identified, is summarized as follows:
\begin{enumerate}

\item The often found correlation between CME speed and SXR peak flux is confirmed. 
\item The correlation of the CME speed is slightly higher with SXR fluence ($r=0.58 \pm 0.09$) than with SXR flux ($r=0.48 \pm 0.12$)
\item The SXR-inferred CME speed performed better than the speed measured by LASCO as an input to the arrival time prediction of ICMEs at Earth using a simple empirical interplanetary propagation model based on \cite{Gop01}.

\end{enumerate}

\subsection{Comparison with Earlier Work}

Detailed comparisons of the kinematical evolution of CMEs in the low corona revealed a close relationship with energy release to the thermal plasma observed in SXR \citep{Zha01,Zha04}. The statistical studies of \cite{Mar07} and \cite{Bei12} demonstrated that the CME acceleration is usually pronounced between the start and peak of the SXR burst, with a maximum near the time of the steepest rise of the time profile. After the SXR peak the CME propagates at roughly constant speed in the corona. This relationship suggests a correlation between the terminal speed of the CME and parameters of the SXR burst, although exceptions from the general trend do exist \citep{Mar07} and are expected to blur the correlation.

The correlation coefficient between the logarithms of CME speed and of SXR peak flux derived in the present work, $r = 0.48 \pm 0.12$, is  similar to values reported by others: 
 $r=0.47$ \citep{Moo02}, $r=0.35$  \citep{Vrs05}, $r=0.50$ \citep{Yas09}, $r=0.32 \pm 0.13$ \citep{Bei12}. A distinctly higher correlation, $r= 0.93$, was found by \cite{Moo03} in a carefully selected small sample of eight flare-CME events, where for four of them, located on the solar disk, the CME speed was corrected for projection effects.

\cite{Moo02}, \cite{Yas09} and the present study were restricted to limb CMEs, where projection effects on the CME speed measurements are expected to be minimized. While the correlation coefficients in these limb event studies are higher, the increase is not significant when compared with the statistical uncertainties derived in the present study and \cite{Bei12}. We note, however, that the coefficient of the logarithm of SXR peak flux $F$ in the linear relationship $\log V_{\rm CME} = a \log F + b$ is higher in our study of limb events ($a=0.20 \pm 0.08$) than in the unrestricted sample of \cite{Bei12} ($a=0.08 \pm 0.03$). 

The correlation is only slightly increased when the SXR fluence is used ($r=0.58 \pm 0.09$) instead of the SXR peak flux. \cite{Yas09} found $r=0.56$ for a larger sample, but without error estimate. So the use of fluence does not seem to significantly improve the correlation between SXRs and CME speed. \cite{Bur04} considered the correlation of the kinetic energy of the CME, instead of the speed, with SXR peak flux of limb events. They reported a high correlation ($r=0.74$ for 24 events), well above the $r=0.48$ of \cite{Yas09}. The absence of an error estimate precludes a comparison of the two values, but the scatter plot in Figure~6 of \cite{Bur04} suggests that the high correlation coefficient is favored by the two extreme events of their sample, and that a lower value might be obtained from a larger sample.

We conclude that the focus on the limb events did provide an improved determination of the relationship between the logarithms of CME speed and of SXR fluence and peak flux. 
But a considerable scatter remains, probably due to physical differences between individual events. In their analysis of a 2-D model of a flux rope eruption, \cite{Ree10} found a power-law relationship between the peak acceleration and the peak SXR flux for a given reconnection rate, measured by the Alfv\'en Mach number of the plasma inflow into the current sheet. The authors showed that for a given CME peak acceleration the peak GOES flux is expected to increase with decreasing reconnection rate, and concluded that different reconnection rates may contribute to explaining the broad scatter in the observed relationships between CME kinematics and SXR emission.

\subsection{SXR Observations and the Prediction of ICME Arrival at the Earth}

We tested the performance of the SXR fluence as a proxy of the CME speed by applying it to the prediction of the ICME arrival near the Earth, using an empirical interplanetary acceleration model based on \cite{Gop01}. For a set of 26 well-defined CME-ICME pairs with associated SXR bursts we found that the SXR-inferred speed tended to perform better than the plane-of-the sky expansion speed measured by a coronagraph on the Earth-Sun line. This suggests that SXR observations can serve as an input to ICME prediction schemes, provided the existence of a CME is ascertained by coronagraphic observations. Problems arise with particularly slow and particularly fast CMEs, where our empirical relationship seems to be a poor predictor. This is probably at least partly due to an inadequate treatment of the bias of the CME-SXR relationship due to the incomplete detection of slow CMEs and faint SXR bursts. Comparisions of selected events with CME speed from STEREO measurements and 3-D modeling confirm the performance of the SXR fluence as a proxy of CME speed.

Recent work using STEREO emphasizes the importance of the interplanetary dynamics of the CME \citep{Kil12,Col13,Mos14} in arrival time predictions, which cannot be captured by a simple empirical model. But when  sophisticated tools such as heliospheric imaging of the Sun-Earth system from a viewpoint away from the Sun-Earth line are not available, the SXR emission can provide valuable constraints for the ICME arrival prediction.

\begin{acks}
This study made extensive use of the CME catalog, generated and maintained at the CDAW Data Center by NASA and The Catholic University of America in cooperation with the Naval Research Laboratory. SOHO is a project of international cooperation between ESA and NASA. GOES data were provided by NOAA and the Solar Data Analysis Center (SDAC) at NASA Goddard Space Flight Center. The data bases of ACE and omniweb (Wind) were also used. The authors thank P.~D\'emoulin for his help with the identification of ICMEs, and G.~Trottet for providing software for the SXR data analysis. The referee is thanked for the detailed reading of the manuscript and comments. C.S-M gratefully acknowledges the financial support of her doctorate studies given by the University of Costa Rica and the Ministry of Science, Technology and Telecommunications of Costa Rica (MICITT) through the National Council of Scientific and Technological Research (CONICIT). This work was partly supported by Centre National d'Etudes Spatiales (CNES).
\end{acks}

%\bibliographystyle{spr-mp-sola}
%\bibliography{ref}

\end{article}
     
\end{document}